\documentclass[twocolumn,prl,a4paper,superscriptaddress]{revtex4}
\usepackage{comment}
\usepackage{color}
\usepackage{amsmath}
\usepackage{amssymb}
\usepackage{graphicx}
\usepackage{braket} 
\usepackage{esint}
\usepackage{adjustbox} 
\usepackage[utf8]{inputenc}

\usepackage[unicode=true,pdfusetitle,
bookmarks=true,bookmarksnumbered=false,bookmarksopen=false,
breaklinks=true,pdfborder={0 0 1},backref=false,colorlinks=true]
{hyperref}
\hypersetup{
	linkcolor=red,urlcolor=blue,citecolor=blue,anchorcolor=blue} 
\usepackage{ulem} 

\usepackage{dcolumn}\usepackage{bm}

\makeatother

\begin{document}

\title{Quantum dynamics of impurities in a Bose-Einstein Condensate}

\author{Javed Akram}

\address{Department of Physics, COMSATS University Islamabad, 44000, Islamabad Pakistan }
\email{javedakram@daad-alumni.de}
%\vspace{10pt}

\date{\today}
\begin{abstract}
We study the quantum dynamics of the two impurities in a trapped quasi-one-dimensional Bose-Einstein Condensate (BEC). We explore the effect of impurity-BEC and impurity-impurity interaction strengths on the dynamics of impurities inside the Bose-Einstein condensate. By studying the auto-correlation function of impurities and the BEC, we analyze and quantify the trapping of impurities inside the BEC. We find out that for the small value of inter-species coupling strength the BEC starts to oscillate inside the trap.  For mild coupling strengths,  attractive and repulsive impurities are captured after a few cycles of oscillation inside the BEC. In the strong interaction strength regime, the to-and-fro motion of impurities suppressed quite fast. Our conclusion indicates that quench dynamics can be a tool for studying impurity BEC interactions or impurity-impurity interactions. Our analysis shows that the generation of Phonon, shock-waves, soliton-train, and self-trapping is strongly dependent on the impurity-BEC coupling coefficient.  
	
\end{abstract} 


%03.75.Lm, 03.75.Nt, 05.30.Jp, 42.25.Fx, 67.85.−d, 67.80.Mg}

\maketitle

\section{Introduction} \label{Sec-1}
Impurities dynamics in a Bose–Einstein condensate (BEC) offer an appealing  platform to investigate profoundly imbalanced multicomponent systems \cite{PhysRevA.85.023623,Hohmann-2015,PhysRevLett.117.055301,PhysRevLett.117.055302}. 
Ultracold atoms provide one to manipulate and examine systems with population imbalance \cite{PhysRevA.78.041403,PhysRevLett.109.235301,PhysRevB.94.054510}, and analogous impurity-impurity induced interactions \cite{Pascal-2018,Lingua_2018,Yan190,PhysRevResearch.2.023154}. The impurity 
interaction with the ultracold atoms offers the possibility for novel applications. For example, in the low-temperature limit, the interaction of a magnetic single impurity with the neighboring electron gas leads to appealing phenomena called Kondo effect \cite{hewson_1993}. Transport of spin-impurity  through a strongly interacting one-dimensional Bose gas reveal fundamental properties of spin transport \cite{PhysRevLett.99.240404,PhysRevLett.103.150601} and spin-charge separation \cite{PhysRevA.77.013607}. 
A spin-impurity can be represented as a qubit that helps to probe the system, where this setup helps to implement impurity as a single atom transistor  \cite{PhysRevLett.93.140408}. A single impurity ion can be used to probe the system of an ultracold atom cloud dynamics \cite{PhysRevLett.105.133202}. The
transfer of quantum information between the atoms can be achieved by two impurity atoms through phonon exchange in a Bose-Einstein condensate  \cite{PhysRevA.71.033605}. Bright shock
waves and gray/dark bisolitons can be generated by quenching the attractive and repulsive interspecies coupling strength of impurity and the BEC respectively \cite{Akram_2016,PhysRevA.93.023606,PhysRevA.93.033610,Akram_2018,Edmonds-2019}.  

\begin{figure}
\begin{center}
	\includegraphics[height=4.0cm,width=8cm]{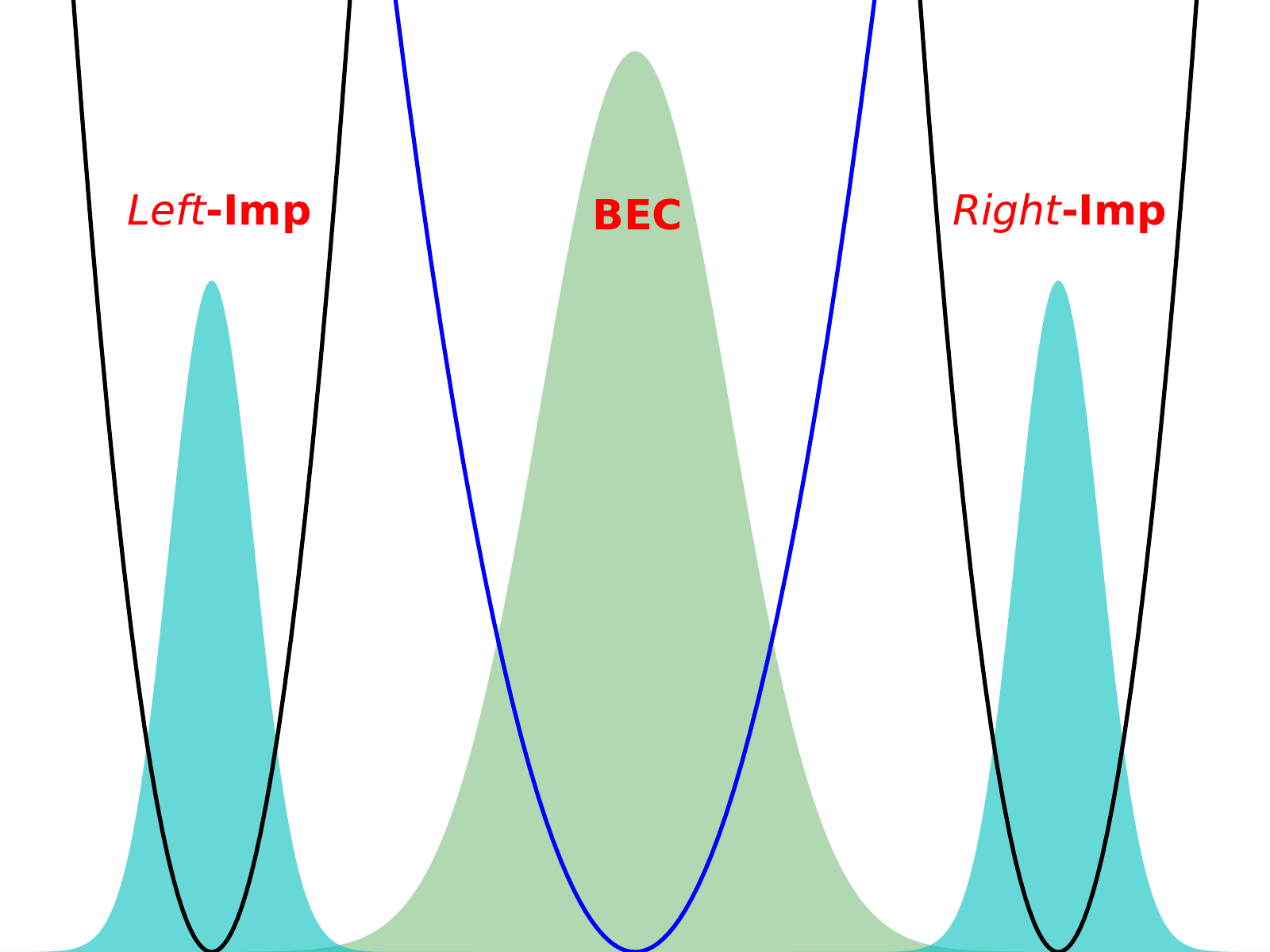}  
	\caption{(Color online) Sketch of the system. The BEC and two impurities initially trapped in separate harmonic traps. At the time $t=0$, impurities traps switched off and let them evolve in the BEC trap. Here, the $Left$-impurity and $Right$-impurity have repulsive ($G_L >0$) and attractive ($G_R <0$) interaction strengths with the BEC, respectively. }
	\label{fig01}
\end{center}
\end{figure}

\begin{figure*}[t]
\begin{center}
\includegraphics[height=10cm,width=15cm]{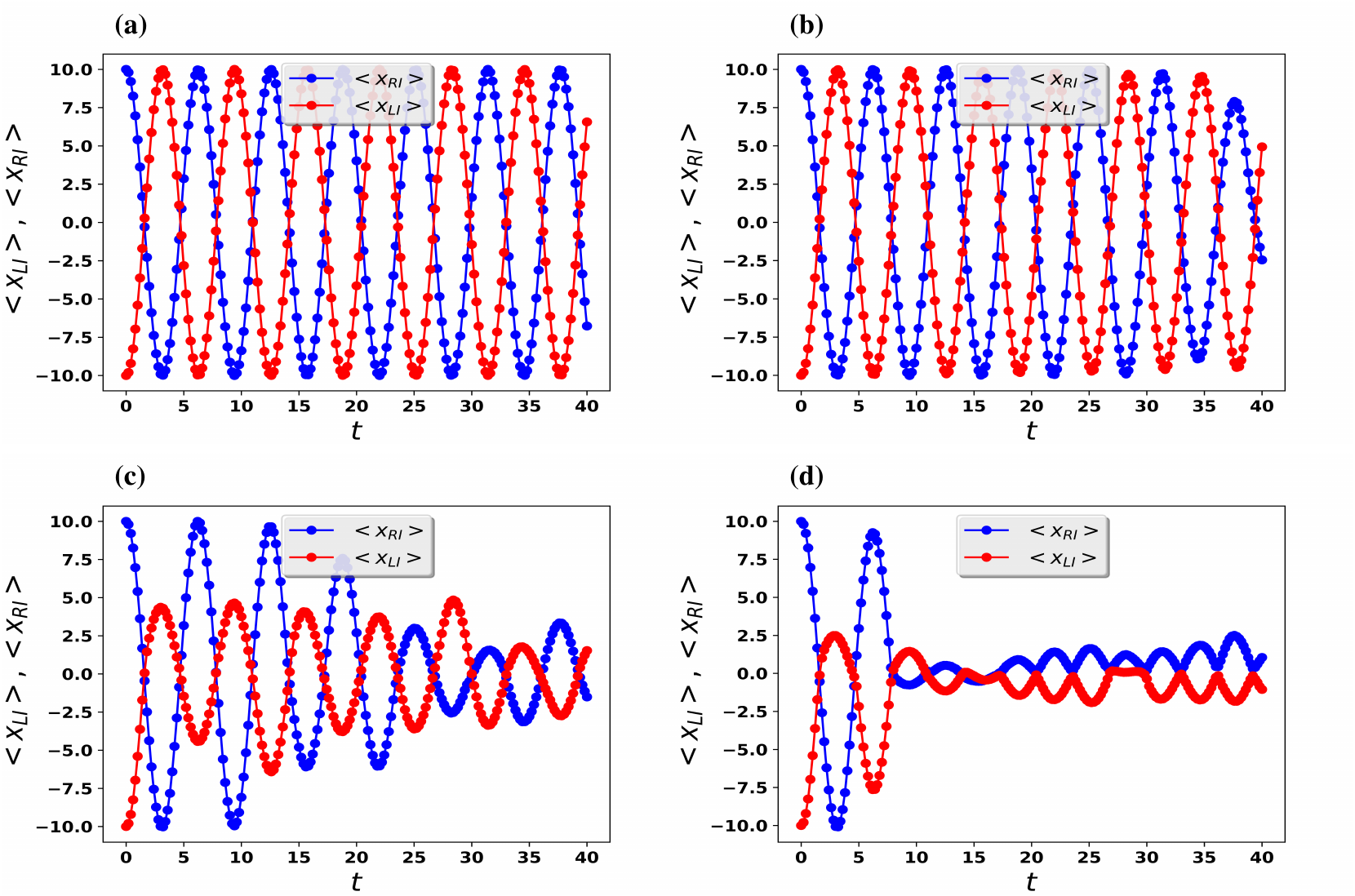}  
\end{center}
\caption{(Color online) We plot mean-position of impurities for (a) $G_{L}=-G_{R}=1$, (b)  $G_{L}=-G_{R}=12$, (c) $G_{L}=-G_{R}=30$ and (d) $G_{L}=-G_{R}=50$. All other dimensionless parameters are $G_B=1$, $G_{II}=0$, and $x_{0R}=-x_{0L}=10$.  }
\label{fig02}
\end{figure*}
In this research, we study the dynamics of the two separated impurity atoms with the BEC, we also discuss the emerging of soliton trains, shock-waves, and self-trapping during the time of flight. By determining the auto-correlation function of impurities, we discuss different regimes, when impurities undergo a state-dependent scattering during the collision with the condensate atoms. This change of states happens due to the generation of phonons. This conditional shift in the density profile of impurities and the BEC is studied, which strongly depends on the interaction strength of impurities and the BEC.  
In addition to this, we also investigate the transmission and reflection of impurities through the BEC for varying interaction strengths. 

In this respect, we organize this paper as follows. We discuss the theoretical model in Sec.~\ref{Sec-2}. A detailed discussion of dynamics of impurities inside the trapped BEC is given in Sec.~\ref{Sec-3}, where this section has two subsections, in one we study the behavior of impurities without impurity-impurity interaction, and in the second we investigate the dynamics of impurities in the presence of the impurity-impurity interaction strength. In Sec.~\ref{Sec-4}, we present the time-of-flight dynamics of impurities and the BEC, where impurities and the BEC traps are switched off and we investigate the possibility of the self-trapping and the generation of soliton-train.  The summary and conclusion are given in Sec.~\ref{Sec-5} and  the last Sec.~\ref{Sec-6} is assigned for acknowledgment.

\section{Theoretical Model}  \label{Sec-2}
The BEC and impurities dynamics in a quasi-1D regime is governed by three coupled Gross-Pitaevskii  equations (GPEs) in dimensionless form, \cite{PhysRevA.58.4836,PhysRevLett.81.5718,PhysRevA.93.033610,Akram_2018}

% \begin{figure}
% \includegraphics[height=8cm,width=9cm]{Fig3.pdf}  
% \caption{(Color online)  We plot width of impurities for (a) $G_{L}=-G_{R}=1$, (b)  $G_{L}=-G_{R}=10$, (c) $G_{L}=-G_{R}=30$ and (d) $G_{L}=-G_{R}=50$. All other dimensionless parameters are $G_B=1$, $G_{II}=0$, and $x_{0R}=-x_{0L}=10$. }
% \label{fig04}
% \end{figure}  

\begin{widetext}

\begin{eqnarray}
i\frac{\partial}{\partial t }\psi_B \left (x,t\right)
 &= \left\{ -\frac{1}{2}\frac{\partial^{2}}{\partial x^{2}}+\frac{x^{2}}{2}
+G_{B}|\psi_B \left (x,t\right)|^{2} +G_{R}  |\psi_{RI}(x,t)|^{2} \right. \\  \nonumber
 & \left.       +G_{L} |\psi_{LI}(x,t) |^{2}\right\} \psi_B\left (x,t\right),\label{Eq03}\\
i\frac{\partial}{\partial t}\psi_{k}\left (x,t\right)&  = \left\{ -\frac{ \alpha^{2}}{2}\frac{\partial^{2}}{\partial x^2}+\frac{V_{I}}{2 \alpha^{2}}+G_{I}|\psi_B (x,t)|^{2} +G_{II}   |\psi_{\bar{k}} (x,t)|^{2}  \right\}    \\  \nonumber
 &   \times  \psi_{k}\left (x,t\right).\label{Eq4}
\end{eqnarray}

\end{widetext}
 
with the normalization condition $\int|\psi_k(x,t)|^{2}dx=1$, here $k\in \{B,RI,LI \}$. Here $\psi_B(x,t)$ and $\psi_{k}(x,t)$ represents the BEC wave function and impurity wave function respectively,  where the subscript  \textquotedblleft $k$\textquotedblright    denotes for the $Left$ and $Right$ side impurities. Additionally, the subscript $\bar{k}$ specify the opposite side of the impurity, e.g., if subscript $k$ describes the $Right$-impurity then the subscript $\bar{k}$  represents the   $Left$-impurity and vice-versa. We define $\alpha=\sqrt{m_{B}/m_{I}}$, here $m_{B}$ describes the mass of the BEC and $m_{I}$ denotes the mass of impurities, in this paper we let the same species for the BEC and for impurities, therefore, $\alpha=1$.  The dimensionless time  defines  as $t$  and the $x$ stands for a dimensionless 1D-space coordinate. For impurities, we let the harmonic confinement as $V_I = \frac{(x-x_{0I})^2}{2}$, where for $Right$-impurity $x_{0R}=10$ and for $Left$-impurity $x_{0L}=-10$.   Here, $G_{B}=2N_{B}\omega_{r}a_{B}/\left(\omega_{x}l_{x} \right)$  represents the dimensionless atom-atom coupling strength for the BEC in a quasi one-dimension, this interaction strength can be tuned by using the the s-wave scattering length, $G_{I} =2a_{IB}\omega_{r}f\left(\omega_{Ir}/\omega_{r}\right)/\omega_{x}l_{x}$ stands for the dimensionless impurity-BEC coupling strength, here, $f\left(\omega_{Ir}/\omega_{r}\right)=\left[1+\left(m_{B}/m_{I}\right)
\right]/\left[1+\left(m_{B}\omega_{r}\right)/\left(m_{I}\omega_{Ir}\right)\right]$
represents a geometric function for more detail please see this Ref. \cite{PhysRevA.93.033610}, which
depends on the ratio of the trap frequencies and $G_{II}$ illustrates the impurity-impurity interaction strength.  The inter-species coupling strength can be tune in experiments by using a number of atoms or by using Feshbach resonances, or by using trap-frequencies ratios \cite{PhysRevLett.93.143001,PhysRevA.79.042718,PhysRevA.95.022715}.    In this paper, we consider single impurities on both ends of the BEC as shown in Fig. \ref{fig01}, therefore, the on side impurity-impurity atoms interaction has not been considered. To make coupled GPEs  (\ref{Eq03},\ref{Eq4}) dimensionless,  we define time in $\omega_{x}^{-1}$, length of the harmonic oscillator depict in $\sqrt{\hbar/{m_B \omega_{x}}}$ and energy characterize in $\hbar \omega_{x}$, here $\omega_{x}$ describes the oscillator length along the $x$-axis.  

To do numerical simulations,  we do discretization of the dimensionless GPEs (\ref{Eq03},\ref{Eq4}), with space step $\Delta{x}=0.0177$ and the time step $\Delta{t}=0.0001$. To perform numerical simulation, we use the time-splitting spectral method \cite{Vudragovic12,Kumar15,Sataric16,LONCAR2016406,YOUNGS2016209}.   To find the ground state wave function of the BEC and impurities, we minimize the energy by simulating in imaginary 
time $\tau=\iota t$, where we let initially the Gaussian wave functions as a test wave function. To achieve the equilibrium, we trap the BEC in a harmonic confinement $V(x)=\frac{x^{2}}{2}$ and impurities in the shifted confinements  $V_I = \frac{(x-x_{0I})^2}{2}$.   
For the dynamical evolution of the wave function of the BEC and impurities, it is eminent to mention that the ground state wave function serves as an initial condition for the rest of the numerical simulations. 
To find out equilibrium wave functions of BEC and impurities, we run the numerical program for separated traps as shown in cartoon Fig. \ref{fig01}. After achieving the equilibrium, we switched off the impurities trap and see the interaction of impurities with the BEC in the one-dimensional confinement $V(x)=x^2/2$. We also investigate another scenario in which all traps are switched off to determine  the time-of-flight dynamics during this free fall.

\begin{figure*}
\begin{center}
\includegraphics[height=10cm,width=15cm]{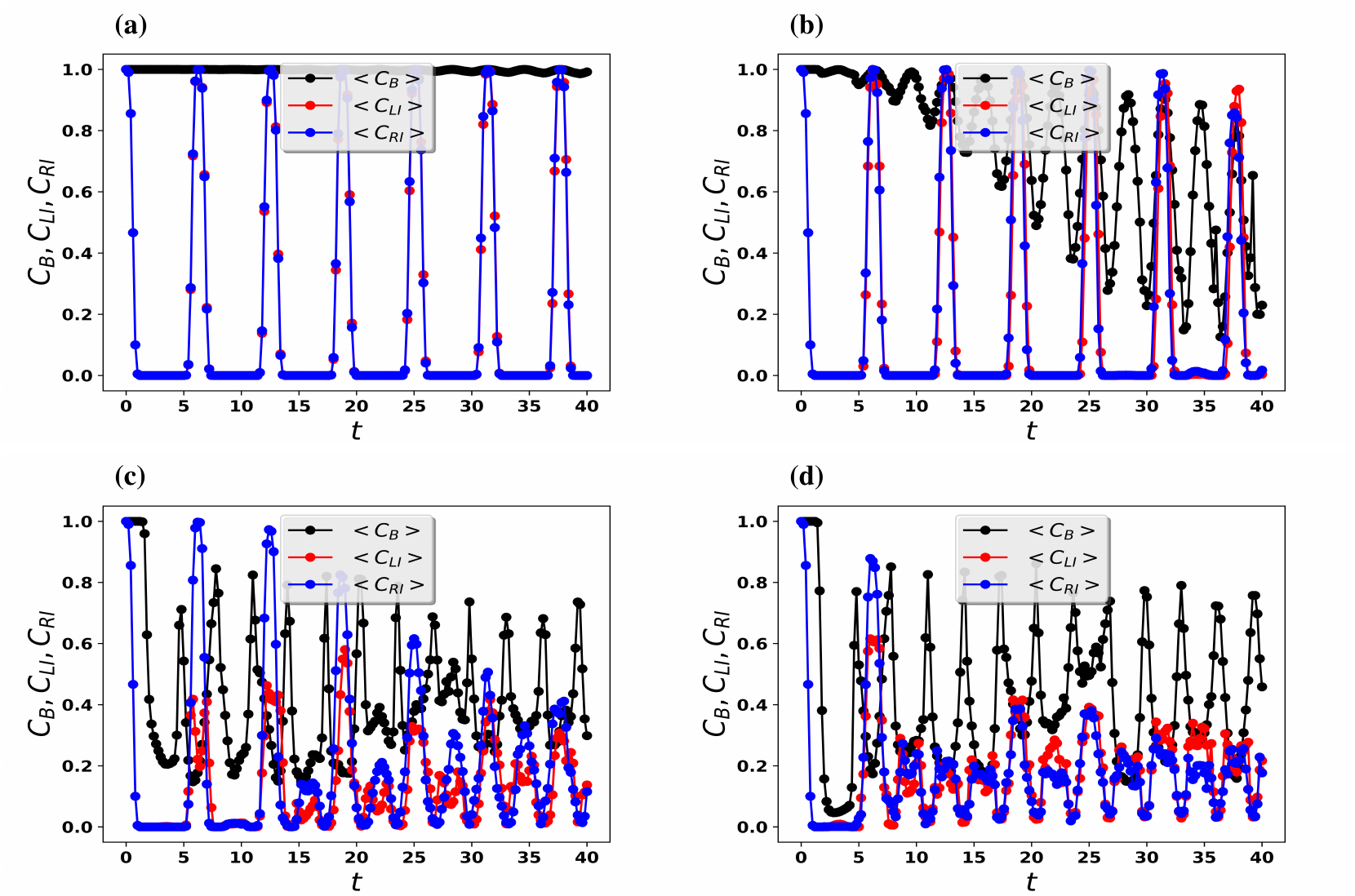}  
\end{center}
\caption{(Color online) We plot auto-correlation of impurities and the BEC for (a) $G_{L}=-G_{R}=1$, (b)  $G_{L}=-G_{R}=12$, (c) $G_{L}=-G_{R}=30$ and (d) $G_{L}=-G_{R}=50$. All other dimensionless parameters are $G_B=1$, $G_{II}=0$, and $x_{0R}=-x_{0L}=10$.    }
\label{fig03}
\end{figure*}

% 
% \begin{figure*}
% \begin{center}
% 	\includegraphics[height=7cm,width=16cm]{fig027.pdf}  
% 	\caption{(Color online) We plot  auto-correlation of impurities and BEC in (a,d), mean-position of impurities in (b,d) and the width of  impurities in (c,f). Here in the upper row and in the lower-row, we have interspecies interaction strength $G_{L}=G_{R}=10$ and $G_{L}=G_{R}=100$,  respectively. All other dimensionless parameters are $G_B=1$, $\bf{G_{II}=-1}$, and $x_{0R}=-x_{0L}=10$. }
% 	\label{Fig8}
% \end{center}
% \end{figure*}

\section{Dynamics of Impurities inside the  trapped BEC} \label{Sec-3}

\subsection{Without impurity-impurity interaction}

Initially, we trap impurities far away from the BEC, the so-called $Left$-impurity ($Right$-impurity) is trapped at $x_{0L}=10 ~ (x_{0R}=-10)$. In our calculation, for the rest of this research article, the inter-species interaction strength between the $Left$-impurity and the BEC is repulsive $G_L>0$ and the inter-species interaction strength between the $Right$-impurity and the BEC is attractive i.e., $G_R<0$.   In the present  scenario, we let the weak BEC coupling strength $G_B=1$ and interaction between the $Right$-impurity and $Left$-impurity is zero i.e., $G_{II}=0$. 
After achieving the equilibrium, we switch off impurities traps and let impurities evolve inside the trapped BEC. For weak BEC-impurity interaction strengths, impurities move to-and-fro motion inside the BEC as shown in Fig. \ref{fig02}(a). Where, the mean-position of impurities is calculated as $<x_{k}>=\int x |\psi_k\left (x,t\right)|^{2}  dx $, here, the subscript $k$ represents the $B$ for BEC, $RI$ for $Right$-impurity, and  $LI$ for $Left$-impurity. To study the dynamics of impurities inside the BEC, we plot in Fig.~\ref{fig03}(a-d)  the auto-correlation function of the BEC and impurities, which we calculate as 
\begin{equation}
 C_{k}= \int \psi_{k}(x,t=0) \psi_{k}(x,t) dx. 
\end{equation}
We note that the auto-correlation function for impurities and the BEC is not affected by low interspecies interaction strengths  $G_{L}=-G_{R}=1$ (Fig.\ref{fig02}(a)), so we can say that impurities preserve their shapes. For the lower values interspecies interaction strengths, the density profile of the BEC is not affected by the movement of impurities, as shown in Fig. \ref{fig03}(a).  We also notice that for mild interspecies interaction strength ($G_{L}=-G_{R}=12$) of impurities and the BEC  mean position is  not too much affected as  presented in Fig.~\ref{fig02}(b), and the BEC dynamics shows the signature of oscillations, therefore, the auto-correlation of the BEC is minimized from its maximum value $one$ during time intervals $t=[ (6,12), (13,17), (20,25), (26,29) ]$ as shown in Fig. \ref{fig03}(b). Additionally, we also note that the auto-correlation function for the $Left$-impurity and $Right$-impurity is also start to decrease from $one$ after dimensionless time $t\approx 30$ as depicted in Fig. \ref{fig03}(b).  
But for strong  BEC-impurity interspecies interaction strength $G_{L}=-G_{R}=30$ and $G_{L}=-G_{R}=50$ the  mean position of impurities is started to localize at the center of the BEC as predicted in Fig. \ref{fig02}(c,d) respectively. For this special case, the BEC start effecting by impurities collisions and quasi-particles (phonons) are generated as impurities are colliding again and again with the BEC atoms during their to-and-fro motion, which leads to the localization of impurities in the center of the trap as depicted in Fig.~\ref{fig02}(c-d). The self-trapping of impurities in the center of the BEC are the results of the nonlinearity of inter- and intra-species interactions strength in GPEs as shown in Fig. \ref{fig03}(c,d).  At the same time, the auto-correlation graph of the BEC and impurities shows that the BEC and impurities lose their shapes due to the strong interaction strength for a long period of time as depicted in Fig. \ref{fig03}(c,d). We also observe that the $Left$-impurity is localized faster at the center of the BEC as compared to the $Right$-impurity as illustrated in Fig. \ref{fig02}(c,d). Therefore, we conclude that for experimentalists the repulsive inter-species impurities are preferable to trap inside the BEC at a short time scale. For intermediate couplings, i.e., around $G_{L}=-G_{R}=30$ and $G_{L}=-G_{R}=50$  regardless of the impurity-BEC coupling strength sign, after a few cycles of oscillation the impurity is trapped by the BEC cloud, and drastically  changes its amplitude of oscillation as shown in Fig.~\ref{fig02}(c-d), this phenomenon is also reported by another study \cite{Lingua_2018}.

\begin{figure*}
\begin{center}
	\includegraphics[height=8cm,width=16cm]{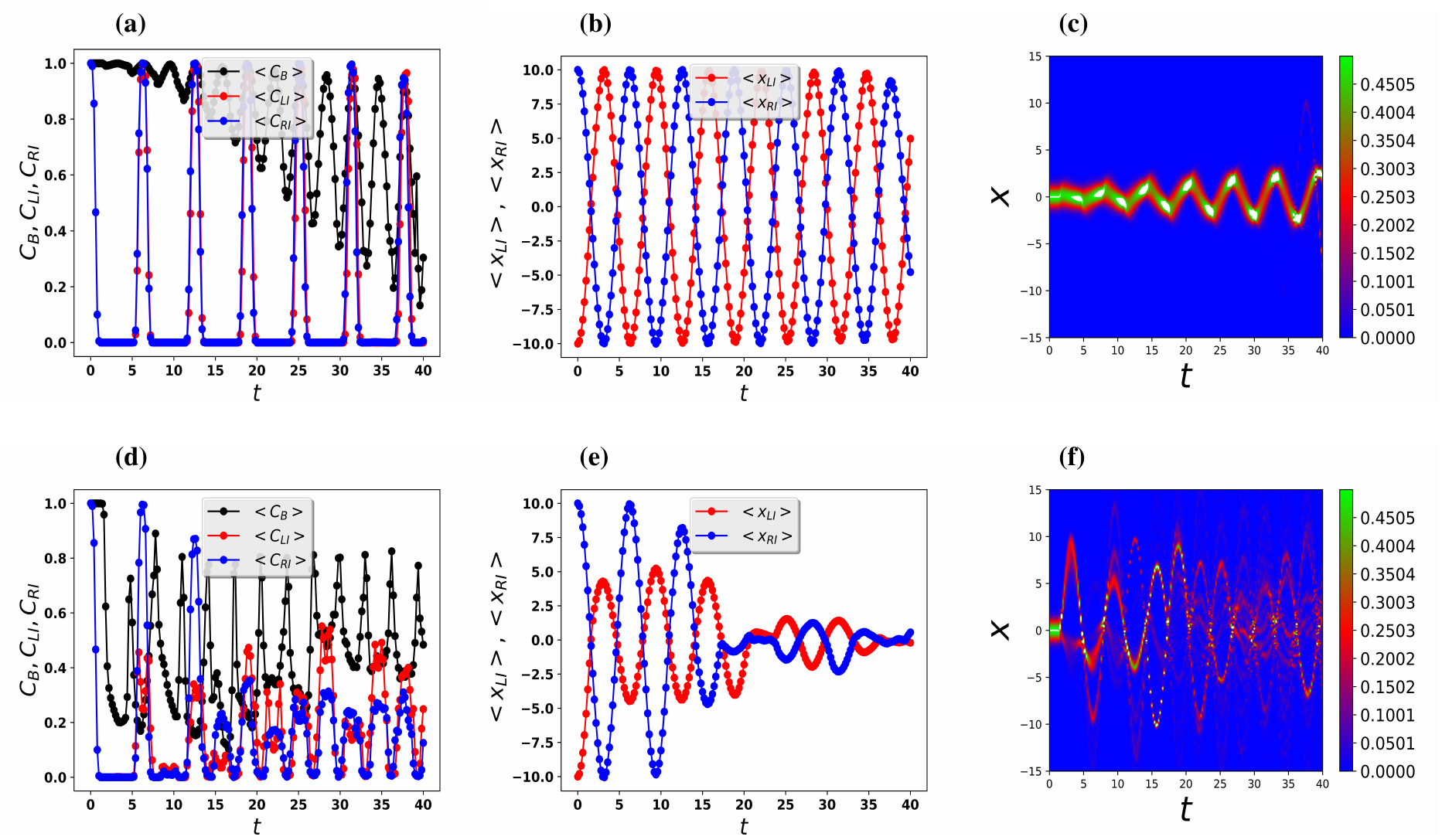}  
	\caption{(Color online) We plot  auto-correlation of impurities and BEC in (a,d), mean-position of impurities in (b,d) and the temporal density of  BEC in (c,f). Here in the upper-row and in the lower-row, we have interspecies interaction strength $G_{L}=-G_{R}=10$ and $G_{L}=-G_{R}=30$ respectively. All other dimensionless parameters are $G_B=1$, $G_{II}=10$, and $x_{0R}=-x_{0L}=10$. }
	\label{fig04}
\end{center}
\end{figure*}

\subsection{With $Right$-impurity and $Left$-impurity interaction}

In this subsection, we discuss the dynamics of impurities in the BEC, during the presence of the mild $Right$-impurity and $Left$-impurity repulsive interaction strength $G_{II}=10$. We would like to study the effect of the repulsive interaction strength on the localization of impurities. We notice that for the weak attracted impurity-impurity interactions and for the small BEC-impurities attractive and repulsive interaction strengths, the dynamics of impurities is not too much different from the  previously discussed cases Fig.~\ref{fig02}, and Fig.~\ref{fig03}, therefore, we do not present these results here. However, a very interesting case happens as we increase the impurity-impurity interaction strength $G_{II}=10$, the auto-correlation of impurities shows that impurities preserved their shape as  shown in Fig.~\ref{fig04}(a) and also the mean position of impurities does not effect by impurity-impurity interaction as depicted in Fig.~\ref{fig04}(b). On the other hand, the BEC temporal density plot shows that the BEC starts oscillating due to the collisions of impurities as can be seen in Fig.~\ref{fig04}(c), this shows the transfer of momentum from impurities to the BEC. For our proposed model, we  consider that the total energy and momentum are conserved, therefore the transfer of momentum decreases the oscillation of impurities as predicted in Fig.~\ref{fig04}(b) for time $t \approx 30$. This situation can be made clearer by increasing the impurity-BEC interaction $G_{L}=-G_{R}=30$  and can be seen in Fig.~\ref{fig04}(e).   This phenomenon helps to localize impurities in the center of the BEC as predicted in Fig.~\ref{fig04}(e), however, this transfer of momentum due to impurities collisions create quasi-particles and solitons in the BEC as shown in Fig.~\ref{fig04}(f) for more detail please see the Refs. \cite{Akram_2016,PhysRevA.93.023606,PhysRevA.93.033610,Akram_2018}.   In addition to that, the shape of the density profile of impurities is also be compromised therefore, the  auto-correlation of impurities is less than one as plotted in Fig.~\ref{fig04}(d).  We also note that the BEC preserves its shapes as its auto-correlation oscillate between $40 \%$ to   $60 \%$. To study explicitly, we consider attractive ($Right$-Impurity) and repulsive ($Left$-Impurity)  interaction strengths of impurities with the BEC, nevertheless, both impurities approximately are localized at the same time scale as shown in Fig.~\ref{fig04}(e). Additionally, our study also reveals that the $Left$-impurity has squeezed oscillation in the BEC  as compared to the $Right$-Impurity, which is due to the repulsive interaction strength of the impurity with the BEC as presented  in Fig.~\ref{fig04}(e). Furthermore, the $Left$-impurity loses its density-profile shape as can be seen from the auto-correlation graph Fig.~\ref{fig04}(d), therefore, it is suggested that the attractive interaction between the BEC and the impurity preferred for the trapping of the impurity inside the BEC.  We also notice that for the weak attractive  impurity-impurity interaction strength, the dynamics of impurities inside the BEC are not so much changed therefore to avoid repeat discussion, we do not present these results here.

\begin{figure*}
\begin{center}
	\includegraphics[height=10.5cm,width=16cm]{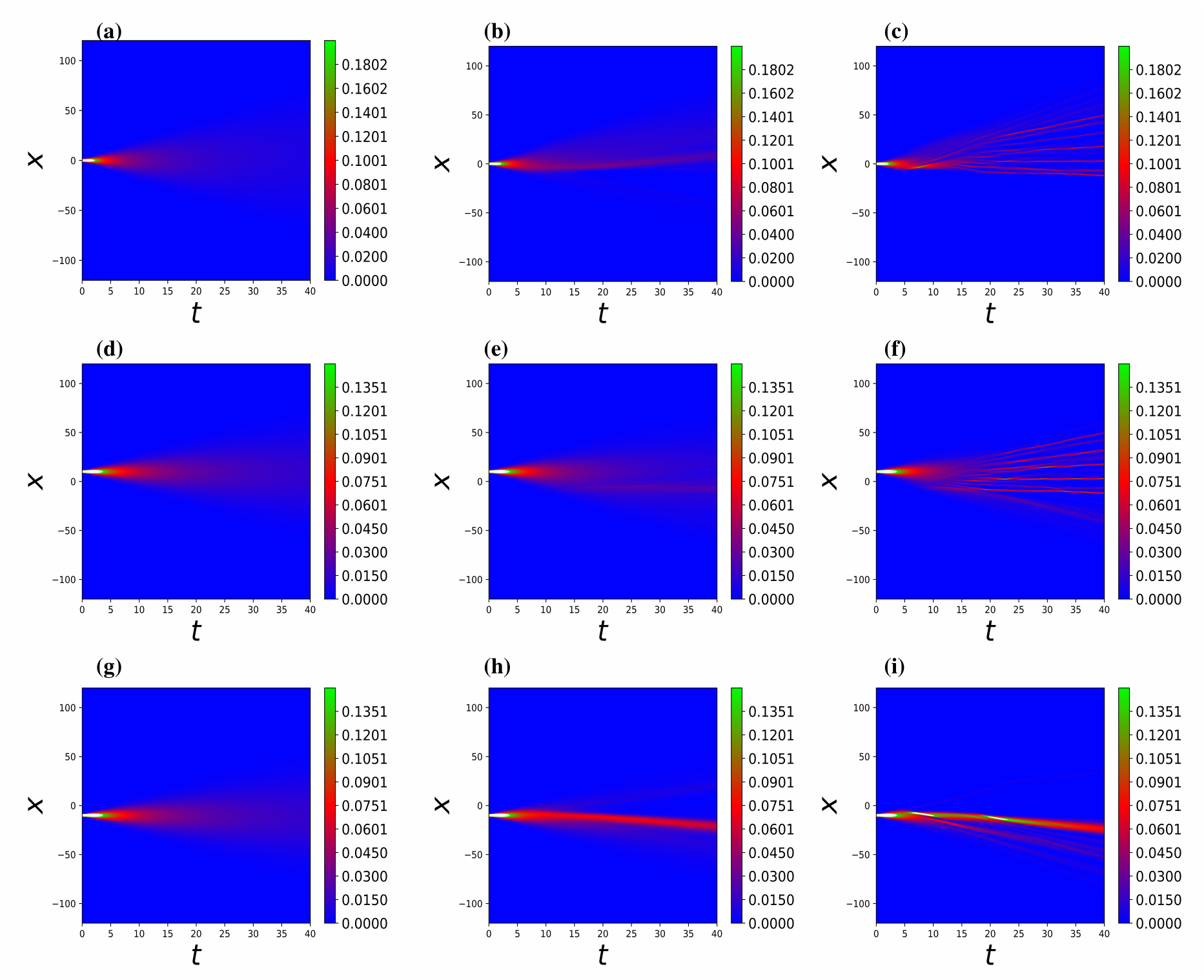}  
	\caption{(Color online) We plot density profile of the BEC (a-c), $Right$-impurity (d-f) and $Left$-impurity (g-i) for different values of impurity-BEC interspecies interaction strength $G_{L}=-G_{R}=1$ (a,d,g), $G_{L}=-G_{R}=10$ (b,e,h) and $G_{L}=-G_{R}=30$ (c,f,i). All other dimensionless parameters are $G_B=1$, $G_{II}=0$, and $x_{0R}=-x_{0L}=10$. }
	\label{fig05}
\end{center}
\end{figure*}

\section{Time-of-flight dynamics of impurities and the BEC} \label{Sec-4}
\begin{figure*}
\begin{center}
	\includegraphics[height=7.5cm,width=16cm]{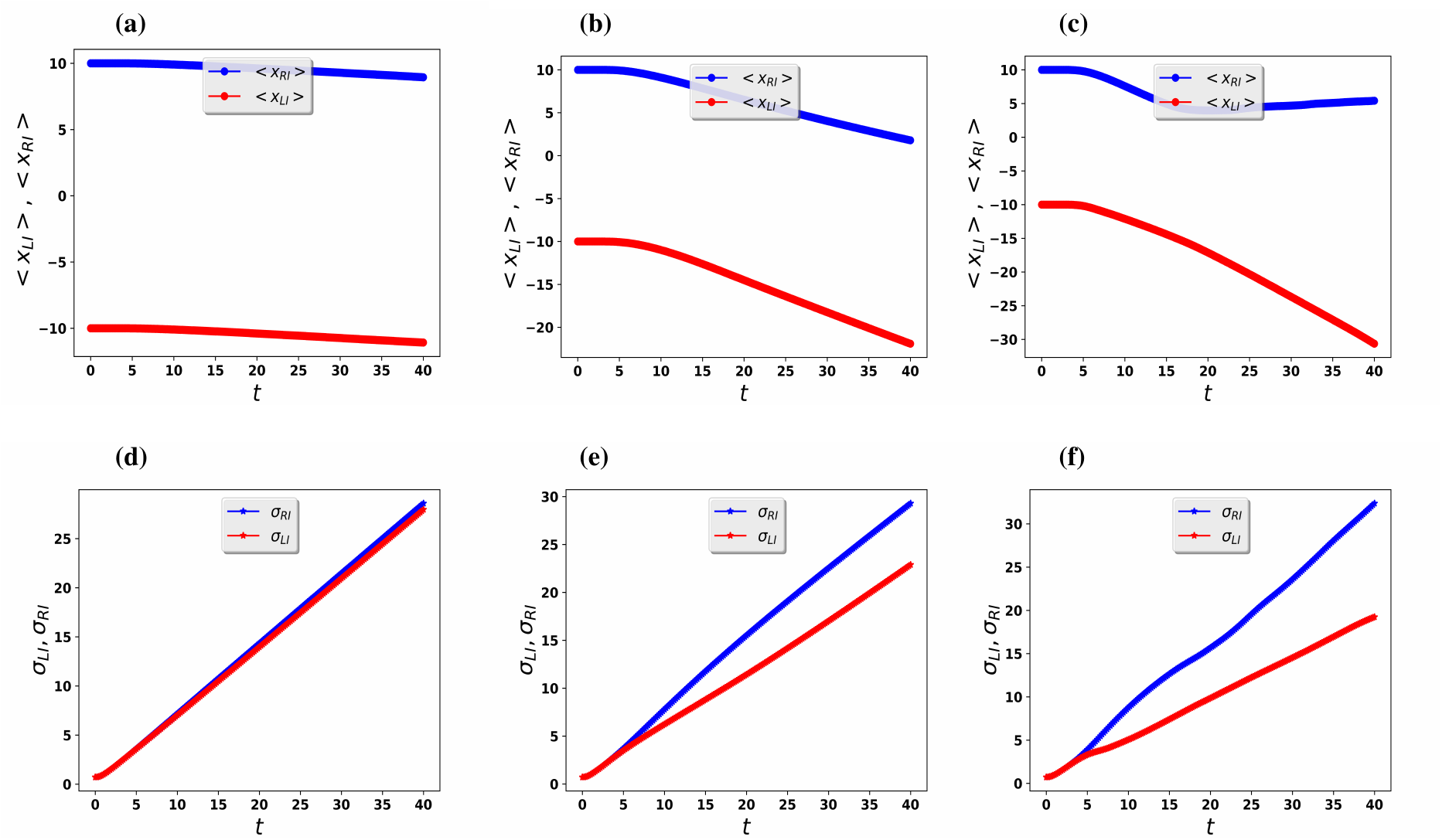}  
	\caption{(Color online) Mean-position (a-c)  and width (d-f) of the of impurities. Here, the impurity-BEC interspecies interaction strength $G_{L}=-G_{R}=1$ (a,d), $G_{L}=-G_{R}=10$ (b,e) and $G_{L}=-G_{R}=30$ (c,f). All other dimensionless parameters are $G_B=1$, $G_{II}=0$, and $x_{0R}=-x_{0L}=10$. }
	\label{fig06}
\end{center}
\end{figure*}

In this section, we discuss the dynamics of the BEC and impurities in the time-of-flight picture. The time-of-flight dynamics of the BEC and impurities can be achieved by switching off the confinement traps of the BEC and impurities as shown in Fig.~\ref{fig05}(a-i). We note that  the $Right$-impurity which has attractive interaction strength between impurity and the BEC is attracted towards the BEC as shown in Fig.~\ref{fig05}(d-f). On the other hand,  the $Left$-impurity which has repulsive interaction strength between impurity and the BEC is squeezed from one side and repealed away from the other side  as shown in Fig.~\ref{fig05}(g-i). The repulsion gets stronger for the strong impurity-BEC interaction strength  as predicted in Fig.~\ref{fig06}(b,c), where we plot the mean-position of the BEC. Hence the BEC temporal density is squeezed from one side and broaden from the other side. In addition to that during the collision of impurities and the BEC, Phonons, and shock waves are generated as can be seen in Fig.~\ref{fig05}(c,f,i). For the strong $G_L=-G_R=30$ interaction strength soliton train is generated as shown in Fig.~\ref{fig05}(c). The $Right$-impurity is broke up into solitonic trains, however, the $Left$-impurity is confined and repealed away from the BEC. We also note that the trapping of BEC fragment between two consecutive soltions as presented in Fig.~\ref{fig05}(c) and Fig.~\ref{fig05}(f).  In addition to that, we also note that the mean-position of impurities is not affected during the time-of-flight for very small impurity-BEC interaction strength $ G_L =  -G_R  = 1$. However, for $ G_L =  -G_R  = 10$,  the mean-position of the impurity is  slightly attractive towards the BEC as shown in Fig.~\ref{fig06}(a-b) and the $Right$-impurity is repelled  away from the BEC center as predicted in Fig.~\ref{fig06}(a-b). In the very strong impurity-BEC interaction regime, for $ G_L =  -G_R  = 30$ a very strange phenomenon happens, firstly, the $Right$-impurity is attractive towards the center of the BEC but after some time  around $t \approx 15$ the impurity repelled away as can be noticed in Fig.~\ref{fig06}(c) which is due to the generation of shock-waves and solitons, which means that the most of the weight of the impurity is pushed away from the BEC. For $Left$-impurity, we notice that the mean-position of the impurity is not affected by strong impurity-BEC interaction strength, and it continuously repealed from the center of the BEC as shown in Fig.~\ref{fig06}(c).  The width of the BEC and impurities grow linearly for small impurity-BEC interaction strength $ G_L =  -G_R  = 1$ as displayed in  Fig.~\ref{fig06}(d), however, for the strong interaction strength  $ G_L =  -G_R  = 10$ the $Right$-impurity width grows linear but the $Left$-impurity width growth is squeezed due to the repulsion of the BEC as presented in  Fig.~\ref{fig06}(e), this phenomena quite visible as shown in  Fig.~\ref{fig05}(i). We also note that for small impurity-BEC interaction strengths the auto-correlation  of the BEC decays exponentially however, for strong impurity-BEC interaction strengths, impurities auto-correlation function decays exponentially to zero, we did not present this result here.
 
\section{Conclusion} \label{Sec-5}
In this research paper, we study the impact of the dynamics of impurities in the presence of the Bose-Einstein condensate. We model our system by using three coupled GPEs, where impurities start from the outside of the BEC with some initial momentum and collide with the BEC. We let our system   be very specific, however, the results are quite general, and different scenarios can be classified depending on the inter-species interaction strengths. Our mean-field results can be beneficial for  future experimental and can also help to develop a theory on impurity-BEC collisions. However, better results can be build by  utilizing more extended
time-dependent methods where quantum and thermal fluctuations are taken into account. We find out that the impurity-BEC interaction strength strongly affects the dynamics of the impurity inside the BEC. We also note that the impurity scattering with the BEC generates Phonons, shock-waves, and soliton trains. Our findings suggest  that the impurity switching from one end of the BEC to the other end depends on system parameters. Additionally, we also note that the self-trapping of BEC fragments depends on system parameters and the impurity-BEC interaction strength.  In the  future, we would like to investigate how the finite size corrections upon the BEC can affect the self-trapping of  impurities \cite{PhysRevA.72.033608,Yukalov-2011,Castellanos-2016}. Additionally, we would like to investigate the generation and dynamics of quantum droplets in the presence of impurities.  
% 
% Newton's law tells us that the gravitational acceleration happens in the same direction as the applied  force. However, in our proposed model,  we notice that the mean position of the $Left$ impurity is affected by the impurity-BEC interaction strength, therefore one finds that impurities are accelerated against the applied force. This kind of phenomena is previously studied in spin-orbit
% coupled (SOC) BEC, and writers called it ``negative mass`` effect \cite{PhysRevLett.118.155301}.
 
 \par 
\par 

\section{Acknowledgment} \label{Sec-6}
Javed Akram thank Hamza Qayyum, Tasawar Abbas and Waqas Masood for the insightful discussion. 
 
% 
% \section{Appendix A}
% 
% Complete solution for the  Mean position of the BEC including  ''non-conservative`` part of the Lagrangian 
%  

%\bibliographystyle{acm}

\section*{References}
%\bibliography{bibBECscattering.bib}
% \bibliography{scattering}

\end{document}